\documentclass[aps,prd,twocolumn,showpacs,groupedaddress,nofootinbib]{revtex4}  
\usepackage{graphicx}  
\usepackage{dcolumn}   
\usepackage{bm}        
\usepackage{amssymb}   

\newcommand{\rb}[1]{\raisebox{1.5ex}[-1.5ex]{#1}}

\begin{document}

\hspace{5.2in} \mbox{Fermilab-Pub-08-007-E}

\title{Search for excited electrons in $p\bar{p}$ collisions at $\sqrt{s} = 1.96$~TeV}
%
\author{V.M.~Abazov$^{36}$}
\author{B.~Abbott$^{76}$}
\author{M.~Abolins$^{66}$}
\author{B.S.~Acharya$^{29}$}
\author{M.~Adams$^{52}$}
\author{T.~Adams$^{50}$}
\author{E.~Aguilo$^{6}$}
\author{S.H.~Ahn$^{31}$}
\author{M.~Ahsan$^{60}$}
\author{G.D.~Alexeev$^{36}$}
% ...

\affiliation{\vspace{0.1 in}(The D\O\ Collaboration)\vspace{0.1 in}}
% ...
\affiliation{$^{6}$University of Alberta, Edmonton, Alberta, Canada,
                Simon Fraser University, Burnaby, British Columbia, Canada,
                York University, Toronto, Ontario, Canada, and
                McGill University, Montreal, Quebec, Canada}
% ...
\affiliation{$^{29}$Tata Institute of Fundamental Research, Mumbai, India}
% ...
\affiliation{$^{31}$Korea Detector Laboratory, Korea University, Seoul, Korea}
% ...
\affiliation{$^{36}$Joint Institute for Nuclear Research, Dubna, Russia}
% ...
\affiliation{$^{50}$Florida State University, Tallahassee, Florida 32306, USA}
% ...
\affiliation{$^{52}$University of Illinois at Chicago,
                Chicago, Illinois 60607, USA}
% ...
\affiliation{$^{60}$Kansas State University, Manhattan, Kansas 66506, USA}
% ...
\affiliation{$^{66}$Michigan State University,
                East Lansing, Michigan 48824, USA}
% ...
\affiliation{$^{76}$University of Oklahoma, Norman, Oklahoma 73019, USA}
% ...
\date{\today}

\begin{abstract}
We present the results of a search for the production of an excited state of the electron,
$e^*$, in proton-antiproton collisions at $\sqrt s = 1.96$~TeV. The data were
collected with the D0 experiment at the Fermilab Tevatron Collider and correspond to an
integrated luminosity of approximately $1$~fb$^{-1}$. We search for $e^*$ in the process
$p \bar p \rightarrow e^* e$, with the $e^*$ subsequently decaying to an electron plus
photon. No excess above the standard model background is observed. Interpreting
our data in the context of a model that describes $e^*$ production by four-fermion
contact interactions and $e^*$ decay via electroweak processes, we set 95\%
C.L. upper limits on the production cross section ranging from $8.9$~fb
to $27$~fb, depending on the mass of the excited electron. Choosing the scale for
contact interactions to be $\Lambda = 1$~TeV, excited electron masses below 756~GeV
are excluded at the 95\% C.L.
\end{abstract}

\maketitle

An open question in particle physics is the cause of the observed mass hierarchy of the quark
and lepton SU(2) doublets in the standard model (SM). One proposed explanation for the three
generations is a compositeness model \cite{compositeness} of the known leptons and quarks.
According to this approach, a quark or a lepton is a bound state of three fermions or of a
fermion and a boson \cite{Terazawa1}. Due to the underlying substructure, compositeness models
imply a large spectrum of excited states. The coupling of excited fermions to ordinary quarks
and leptons, resulting from novel strong interactions, can be described by contact
interactions (CI) with the effective four-fermion Lagrangian \cite{Baur90}
$$
 \mathcal{L}_{\mathrm{CI}} \, = \, \frac{g^2}{2\Lambda^2}\, j^{\mu}\, j_{\mu},
$$
where $\Lambda$ is the compositeness scale and $j_{\mu}$ is the fermion current
\begin{eqnarray*}
  j_{\mu} & = &  \eta_L \, \bar{f}_L\gamma_{\mu}f_L +
                 \eta'_L \, \bar{f}^*_L\gamma_{\mu}f^*_L +
                 \eta''_L \, \bar{f}^*_L \gamma_{\mu}f_L \\
  & & + \, h.c. \, + \, (L \rightarrow R).
\end{eqnarray*}
The SM and excited fermions are denoted by $f$ and $f^*$, respectively; $g^2$ is
chosen to be $4 \pi$, the $\eta$ factors for the left-handed currents are
conventionally set to one, and the right-handed currents are set to zero.

Gauge mediated transitions between ordinary and excited fermions can be described by the
effective Lagrangian \cite{Baur90}
\begin{eqnarray*}
  \mathcal{L}_{\mathrm{EW}} & & = \ \frac{1}{2\Lambda} \bar{f}_R^*\,\sigma^{\mu\nu} \\
     & & \left[ g_s f_s \frac{\lambda^a}{2} G^a_{\mu\nu} + g f \frac{\tau}{2} W_{\mu\nu} 
                    +   g'f'\frac{Y}{2}B_{\mu\nu}\right] f_L + h.c.
\end{eqnarray*}
where $G^a_{\mu\nu}$, $W_{\mu\nu}$, and $B_{\mu\nu}$ are the field strength tensors
of the gluon, the SU(2) and U(1) gauge fields, respectively, and $f_s$, $f$ and $f'$
are parameters of order one.

\begin{figure}[ht]
\begin{picture}(245,105)(8,0)
    \put(90,-12) {\includegraphics[scale=0.30]{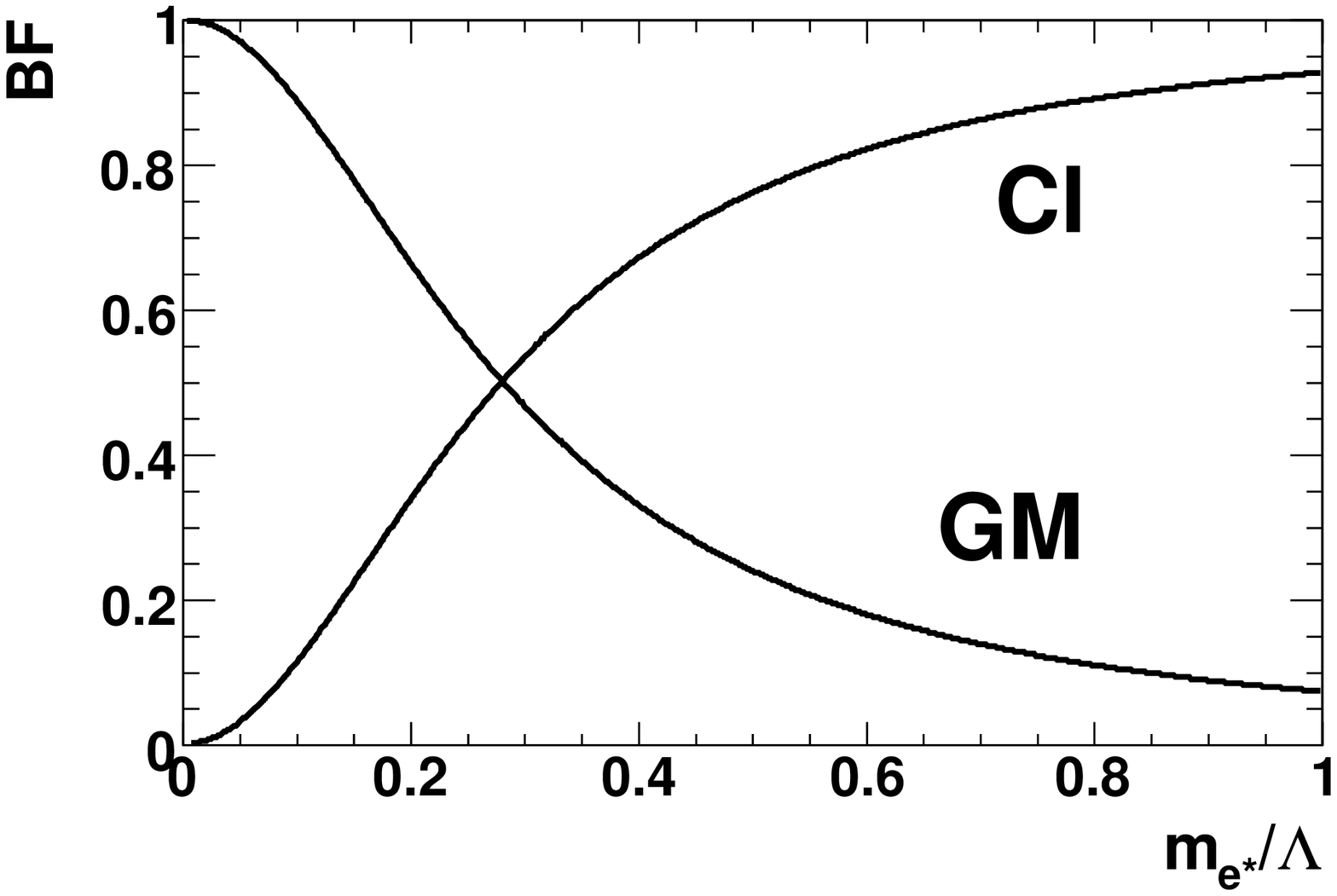}}
    \put(7,20) {\includegraphics[scale=0.42]{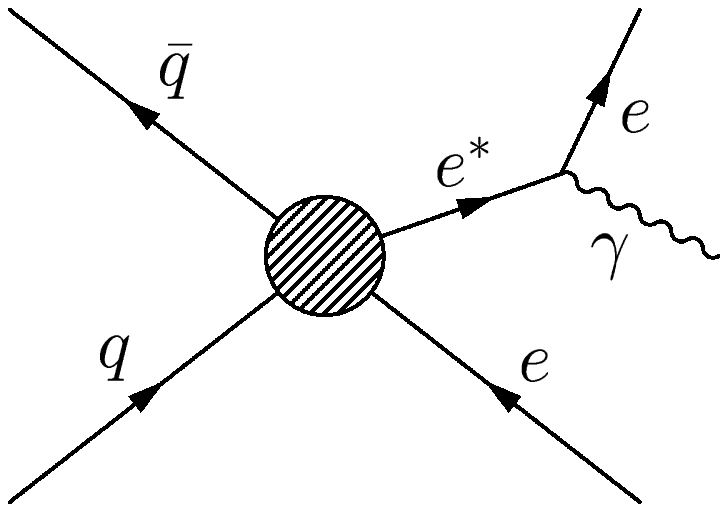}}
    \put(8,104)  {\sf (a)}
    \put(115,104)  {\sf (b)}
    \end{picture}
  \caption{(a) Four-fermion contact interaction $q \bar{q} \rightarrow e^* e$, and electroweak decay
           $e^* \rightarrow e \gamma$. (b) Relative contributions of decays via
	   contact interactions and via electroweak interactions (GM) as a function of $m_{e^*} / \Lambda$.}
  \label{feynman}
\end{figure}

For the present analysis, we consider single production of an excited electron $e^*$ in
association with an electron via four-fermion contact interactions, with the subsequent
electroweak decay of the $e^*$ into an electron and a photon (Fig.~\ref{feynman}(a)). This
decay mode leads to the fully reconstructable and almost background-free final state $e e
\gamma$. With the data considered herein, collected with the D0 detector at the Fermilab
Tevatron Collider in $p\bar{p}$ collisions at $\sqrt s = 1.96$~TeV, the largest expected SM
background is from the Drell-Yan (DY) process $p\bar{p}\rightarrow Z/\gamma^* \rightarrow
e^+e^- (\gamma)$, with the final state photon radiated by either a parton in the initial
state or from one of the final state electrons. This background can be strongly suppressed
by the application of suitable selection criteria. Other backgrounds are small.

Previous searches have found no evidence for the production of excited electrons, e.g.~at
the CERN LEP $e^+e^-$ \cite{LEP} and the DESY HERA $ep$ \cite{HERA} colliders, in the
context of models where the production of excited electrons proceeds via gauge
interactions; however, the reach has been limited by the available center-of-mass energy
to $m_{e^*} \lesssim 300$~GeV. Searches for quark-lepton compositeness via deviations from the
Drell-Yan cross section at the Tevatron have excluded values of $\Lambda$ of up to
$\approx 6$~TeV depending on the chirality \cite{comprun1}. The present analysis is
complementary to those results in the sense that an exclusive channel and different
couplings ($\eta$ factors) are probed. The CDF collaboration has recently presented
results \cite{cdf} for the production of excited electrons which will be discussed later.

For the simulation of the signal the {\sc pythia} event generator \cite{pythia} is used,
following the model of Ref.~\cite{Baur90}. The branching fraction for the decay $e^*
\rightarrow e \gamma$ normalized to all gauge particle decay modes is 30\% for masses above
300 GeV; for smaller $e^*$ masses it increases up to 73\% at $m_{e^*} = 100$~GeV. Decays via
contact interactions, not implemented in {\sc pythia}, contribute between a few percent of all
decays for $\Lambda \gg m_{e^*}$ and 92\% for $\Lambda = m_{e^*}$ \cite{Baur90} (see
Fig.~\ref{feynman}(b)). This is taken into account for the signal expectation. The leading
order cross section calculated with {\sc pythia} is corrected to next-to-next-to-leading order
(NNLO) using Ref.~\cite{dy}; the corresponding correction factor varies between 1.37 and 1.42,
depending on the invariant mass of the electron and the excited electron. The total width is
greater than 1~GeV for $100$~GeV~$\leq m_{e^*} \leq 1000$~GeV, thus lifetime effects can be
neglected. For the values of $m_{e^*}$ and $\Lambda$ studied here, the total width is always
less than 10\% of $m_{e^*}$ \cite{Baur90}.

The dominant SM background process at all stages of the selection is DY production of
$e^+e^-$ pairs. This background, as well as diboson ($WW$, $WZ$, $ZZ$) production, is
simulated with the {\sc pythia} Monte Carlo (MC) program. The DY expectation (as well as
$W\to e\nu$) is corrected using the NNLO calculation from Ref.~\cite{dy}. For diboson
production, the next-to-leading order cross sections from Ref.~\cite{mcfm} are used.
Contributions from $t\bar{t}$ \cite{ttbar} and $W$ boson production are found to be
negligible. Monte Carlo events, both for SM and signal, are passed through a detector
simulation based on the {\sc geant} \cite{geant} package and reconstructed using the same
reconstruction program as the data. The CTEQ6L1 parton distribution functions (PDFs)
\cite{cteq} are used for the generation of all MC samples.

The analysis is based on the data collected with the D0 detector \cite{run2det} between
August 2002 and February 2006, corresponding to an integrated luminosity of $1.01 \pm
0.06$~fb$^{-1}$. The D0 detector includes a central tracking system, which comprises a
silicon microstrip tracker and a central fiber tracker, both located within a 2~T
superconducting solenoidal magnet, and optimized for tracking and vertexing capability
at pseudorapidities\footnote{The pseudorapidity $\eta$ is defined as $\eta = - \ln
\left[\tan(\theta / 2)\right]$. We use the polar angle $\theta$ relative to the proton
beam direction, and $\phi$ is the azimuthal angle, all measured with respect to the
geometric center of the detector.} $|\eta|<2.5$. Three liquid argon and uranium
calorimeters provide coverage out to $|\eta|\approx 4.2$: a central section (CC)
covering $|\eta| \lesssim 1.1$, and two end calorimeters (EC). The electromagnetic
section of the calorimeter has four longitudinal layers and transverse segmentation of
$0.1 \times 0.1$ in $\eta - \phi$ space, except in the third layer, where it is $0.05
\times 0.05$. A muon system resides beyond the calorimetry, and consists of layers of
tracking detectors and scintillation trigger counters before and after 1.8~T iron
toroids. Luminosity is measured using scintillator arrays located in front of the EC
cryostats, covering $2.7 < |\eta| < 4.4$. A three-level trigger system uses information
from tracking, calorimetry, and muon systems to reduce the $p\bar{p}$ bunch crossing
rate of 1.5~MHz to $\approx 100$~Hz, which is written to tape.

Efficiencies for electron and photon identification and track reconstruction are
determined from the simulation. To verify the simulation and to estimate systematic
uncertainties, the efficiencies are also calculated from data samples, using $Z
\rightarrow e^+e^-$ candidate events and other dilepton events for electrons and tracks.
Small differences between the efficiency determinations from data and simulation are
corrected in the simulation. We assume that the different response for electrons and
photons in the calorimeter is properly modeled by the simulation. The transverse (with
respect to the beam axis) momentum resolution of the central tracker and the energy
resolution of the electromagnetic calorimeter are tuned in the simulation to reproduce
the resolutions observed in the data using $Z\rightarrow \ell \ell$ ($\ell = e, \mu$)
events.

The process $p \bar p \rightarrow e^* e$ with $e^* \rightarrow e \gamma$ leads to a
final state with two highly energetic isolated electrons and a photon. First, the two
electrons are identified as clusters of calorimeter energy with characteristic
longitudinal and transverse shower shapes and at least 90\% of the energy deposited in
the electromagnetic section of the calorimeter. Two electrons, with transverse energies
$E_T > 25$~GeV and $E_T > 15$~GeV, are required. Both electrons are matched to tracks
in the central tracking system, and we distinguish between CC ($|\eta|<1.1$ with
respect to the detector center) and EC ($1.5<|\eta|<2.5$) electrons. Events with the
two electrons in opposite EC are rejected in order to suppress multijet background. The
signal is expected to produce isolated electrons, therefore both electrons need to
fulfill ${\cal I} \equiv (E_{\text{tot}}(0.4) - E_{\text{em}}(0.2)) /
E_{\text{em}}(0.2) < 0.2$, where $E_{\text{tot}}(0.4)$ and $E_{\text{em}}(0.2)$ denote
the energies deposited in the calorimeter and deposited in only its electromagnetic
section, respectively, in cones of size $\Delta {\cal R} = \sqrt{(\Delta\eta)^2 +
(\Delta\phi)^2} = 0.4$ and 0.2. The electrons are required to be separated by $\Delta
{\cal R}>0.4$.

\begin{figure} \setlength{\unitlength}{1cm}
  \begin{picture}(8.6,4.4)(0.0,0.0)
    \put(-0.1,-0.4) {\includegraphics[scale=0.227]{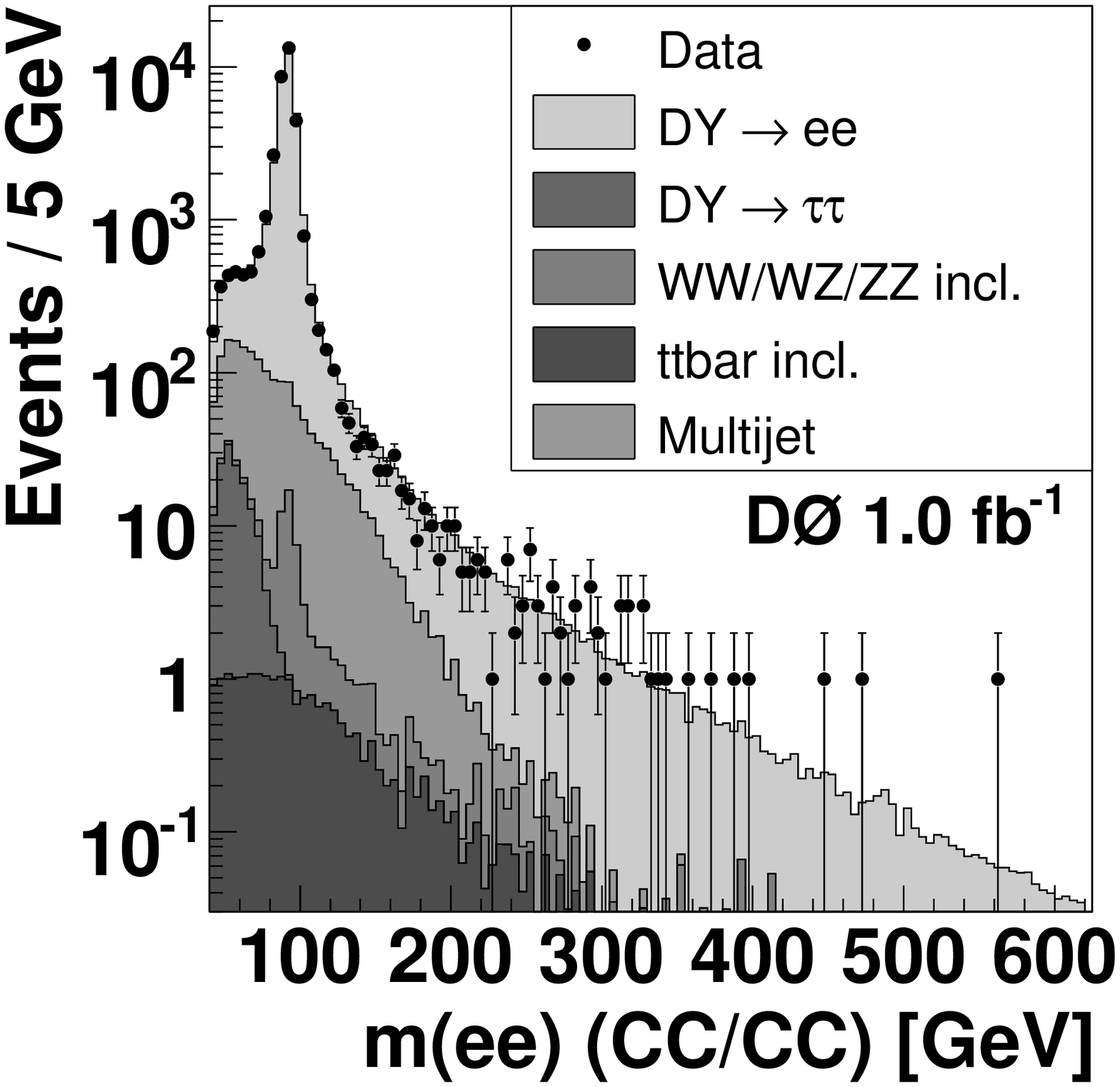}}
    \put(4.3,-0.4)  {\includegraphics[scale=0.227]{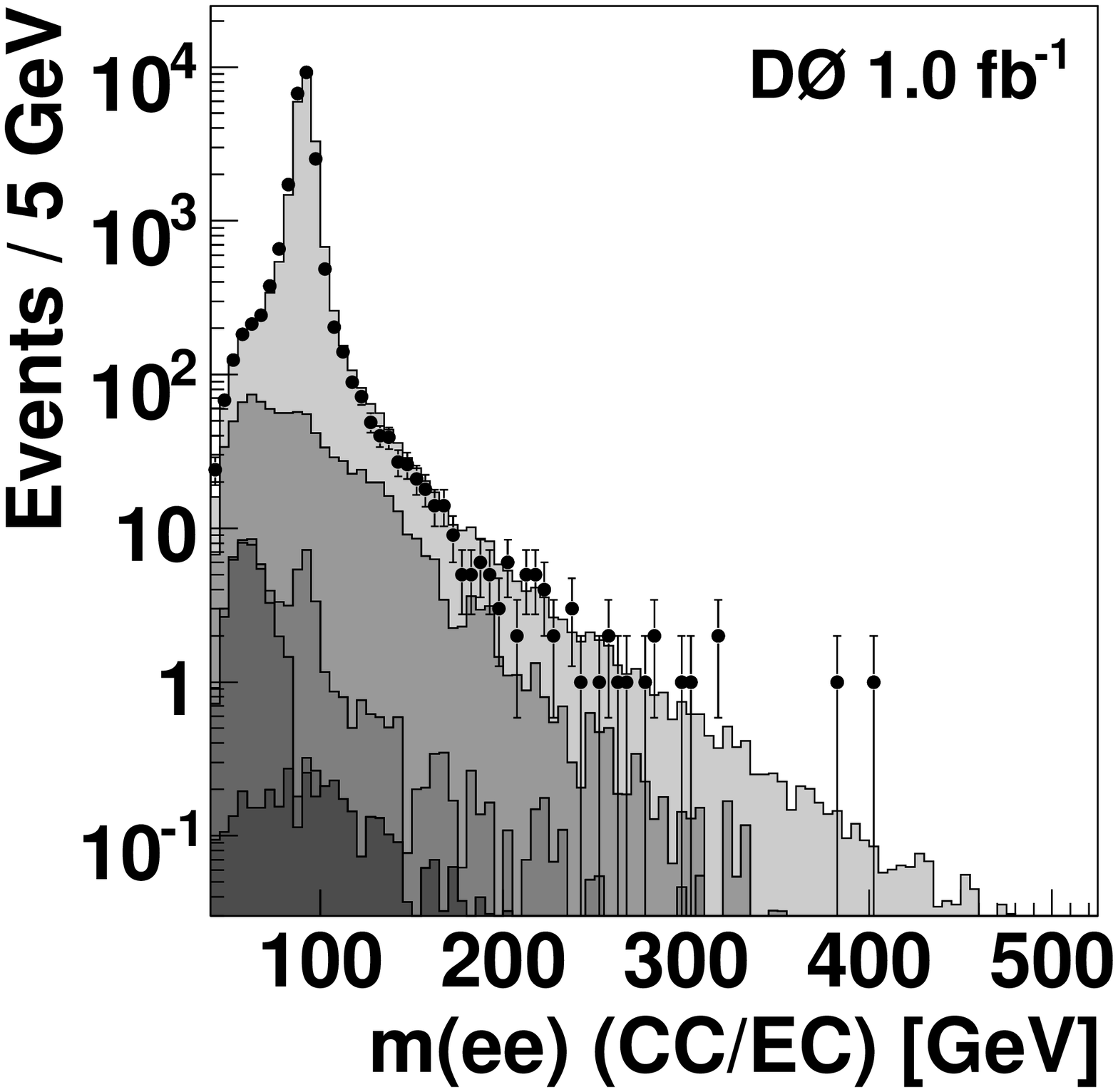}}
    \put(0.8,4.1)  {\sf (a)}
    \put(5.2,4.1)  {\sf (b)}
  \end{picture}
  \caption{Invariant dielectron mass distribution in the dielectron data sample compared to the SM
           expectation (a) for events with both electrons reconstructed in the CC and (b) for
	   events in the CC/EC topology, for data (points with statistical uncertainties) and SM
	   backgrounds (DY, diboson, $t\bar{t}$, and multijet production).}
  \label{control1}
\end{figure}

The events were collected with trigger conditions requiring one or two electrons detected in
the calorimeters, with varying conditions depending on the $E_T$ thresholds, the shower
shape, tracks in the central tracking system, and number of electrons. The overall trigger
efficiency is determined from independent data samples and is consistent with 100\% for the
signal after application of all selection criteria. The selected dielectron sample contains
$62930$ events, whereas $61900 \pm 5700$ events are expected from SM processes. The
invariant dielectron mass distributions for CC/CC and CC/EC topologies are shown in
Fig.~\ref{control1}. The largest SM contribution is DY production of $e^+e^-$ pairs,
followed by multijet production with misidentified electrons. The multijet background is
estimated using a data sample where at least one of the electron candidates fails the shower
shape requirements. This sample is then corrected as a function of $E_T$ and $\eta$ of the
misidentified electrons in order to account for different misidentification rates in the CC
and EC and the different trigger efficiency for misidentified electrons.

Next, a photon is identified in the event as an isolated cluster of calorimeter energy with
at least 97\% of its energy deposited in the electromagnetic section of the calorimeter (CC
or EC). The isolation condition is ${\cal I}  < 0.07$. The photon candidate $E_T$ must be
larger than 15~GeV, no track is allowed to be matched to the photon candidate in $\eta$ and
$\phi$ with a $\chi^2$ probability of greater than $0.1$\%, and the sum of the transverse
momenta of tracks within a hollow cone defined by $0.05 < \Delta{R} < 0.4$ around the photon
direction has to be below 2~GeV to further ensure isolation. Additional shower shape
criteria are imposed to increase the photon purity. The photon candidate is required to be
separated from the electron candidates in the event by $\Delta {\cal R} > 0.4$.

\begin{figure} \setlength{\unitlength}{1cm}
  \begin{picture}(8.6,8.8)(0.0,0.0)
    \put(-0.1,4.1) {\includegraphics[scale=0.224]{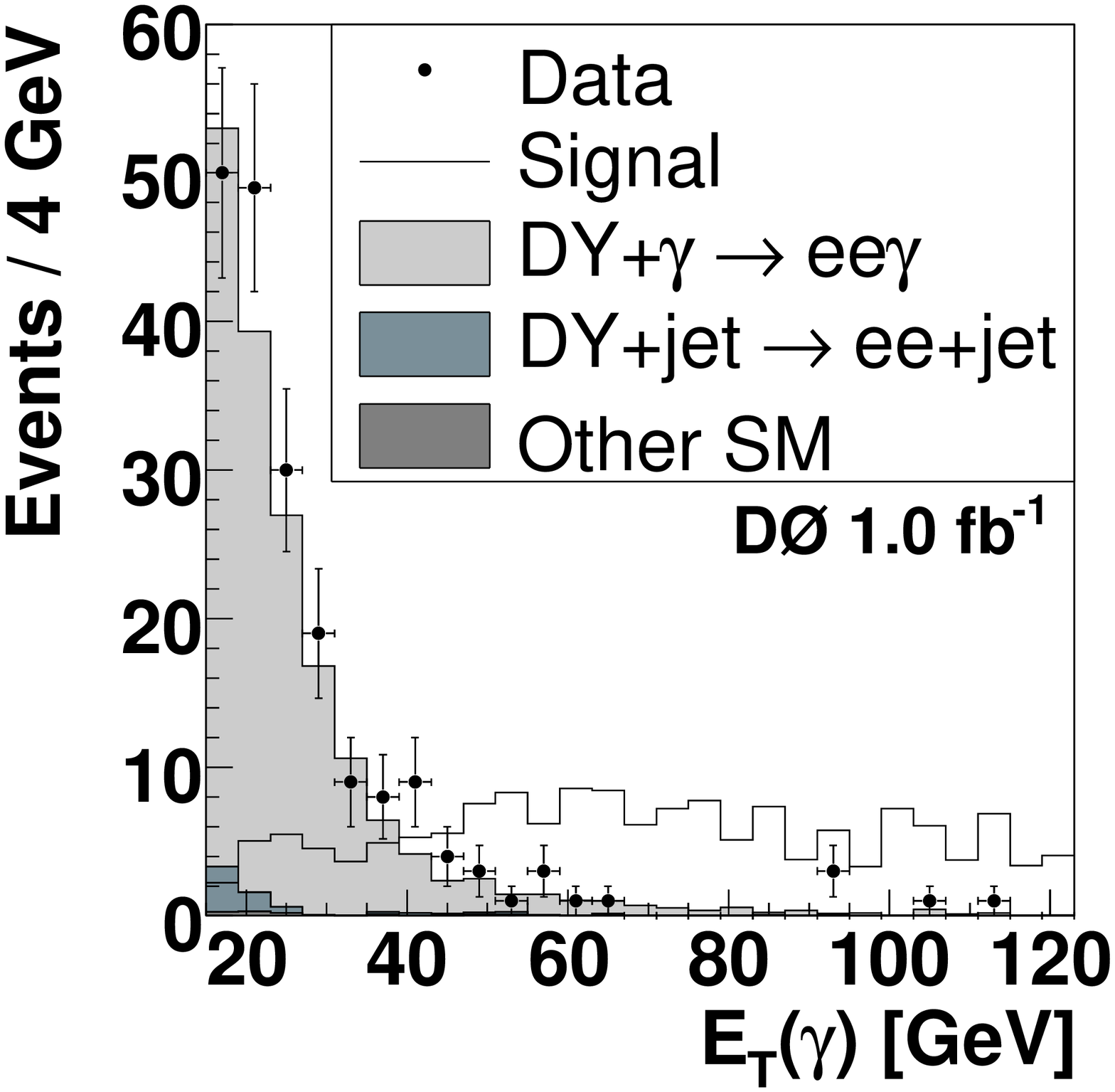}}
    \put(4.3,4.1) {\includegraphics[scale=0.224]{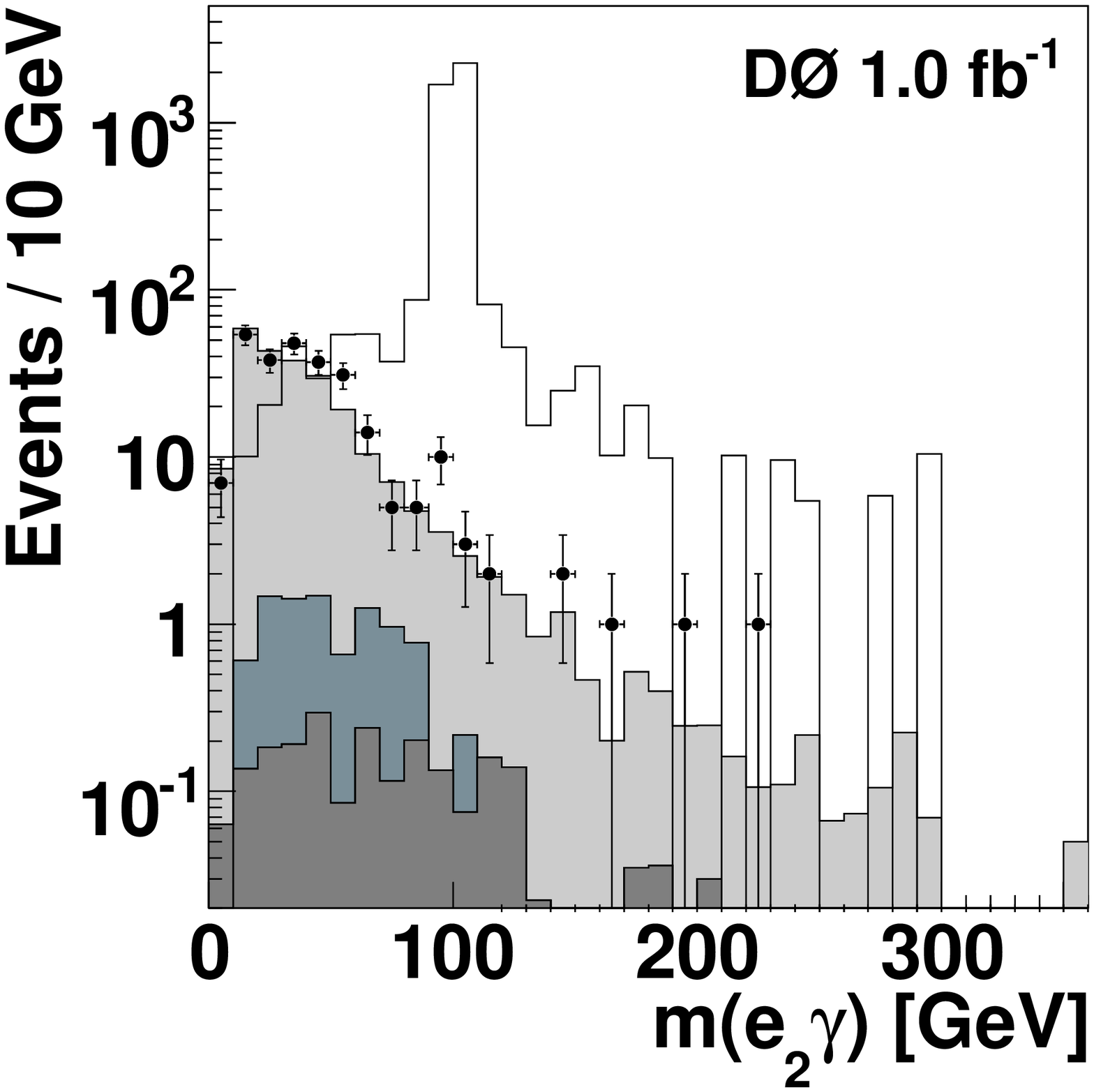}}
    \put(-0.1,-0.4) {\includegraphics[scale=0.224]{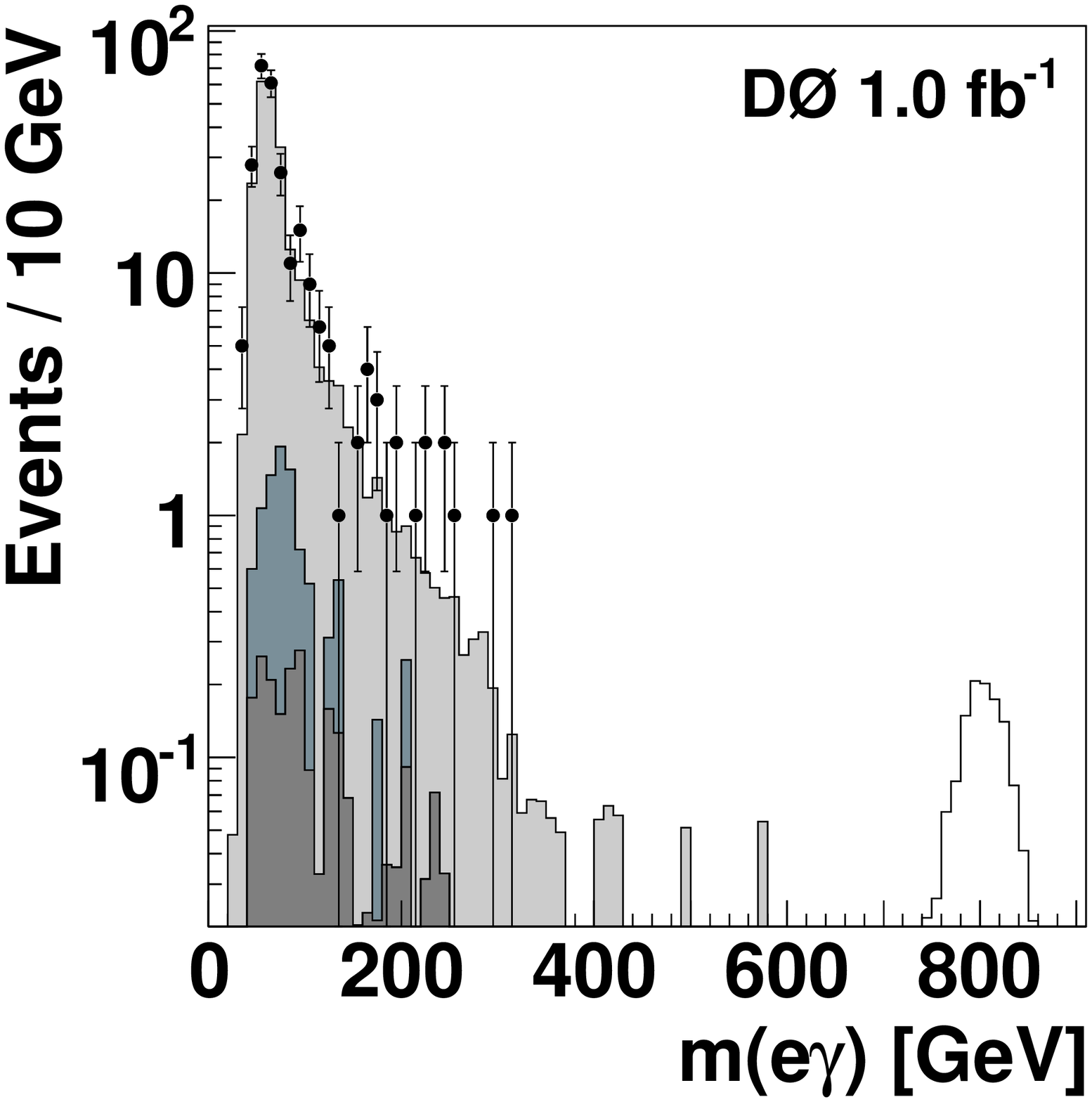}}
    \put(4.3,-0.4) {\includegraphics[scale=0.224]{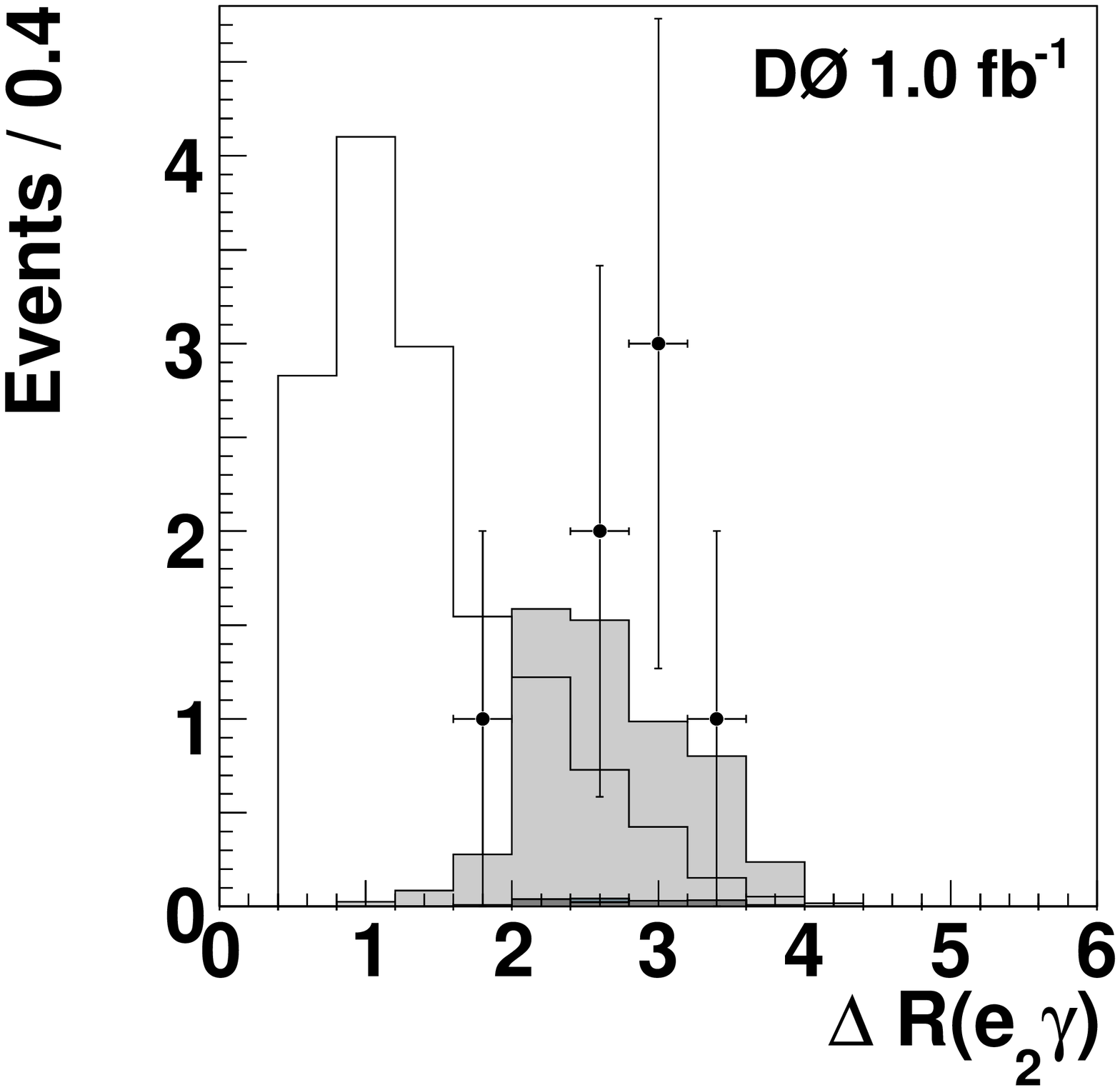}}
    \put(0.8,8.5)  {\sf (a)}
    \put(5.2,8.5)  {\sf (b)}
    \put(0.8,4.0)  {\sf (c)}
    \put(5.2,4.0)  {\sf (d)}
  \end{picture}
  \caption{For the $ee\gamma$ sample, (a) the photon $E_T$ distribution, (b) the distribution
           of the $e_2 \gamma$ invariant mass compared with the SM expectation and a possible
	   $e^*$ signal for $m_{e^*} = 100$~GeV; (c) the $e \gamma$ invariant mass for the $e
	   \gamma$ combination closest to $m_{e^*} = 800$~GeV. In (d) the separation $\Delta
	   R (e_2, \gamma)$ is shown after the cut on the invariant mass $m(e_2,\gamma) >
	   90$~GeV for $m_{e^*} = 100$~GeV. The signal corresponds to $\Lambda = 2$~TeV,
	   1~TeV, 1~TeV and 4~TeV in (a), (b), (c) and (d), respectively. All uncertainties
	   are statistical only.}
  \label{control2}
\end{figure}

After this selection, $239 \pm 36$ events are expected from SM processes. Of these, 226
events are due to DY $\rightarrow e^+e^-$ with a genuine high $E_T$ photon, followed by 7
events from DY $\rightarrow e^+e^- +$ jets, where a jet is misidentified as a photon. The
absolute rate of the latter process has been determined from a data sample enriched in
``fake'' photons, applying the rate for such objects to be misidentified as photons as a
function of $E_T$, and subtracting the true photon contribution \cite{Zgamma}. Finally,
about 4 and 2 events are expected from multijet and diboson production, respectively. In the
data, 259 events are selected, compatible with the SM prediction. The photon $E_T$
distributions for the data and SM background are shown in Fig.~\ref{control2}(a).

Additional selection criteria depending on the hypothetical $e^*$ mass are applied to reduce
the remaining SM background. The following criteria have all been optimized to achieve the
best expected upper limit on the production cross section. The $e^*$ invariant mass can be
reconstructed from one of the electrons and the photon. For $m_{e^*} < 300$~GeV, the lower
$E_T$ electron ($e_2$) is predominantly the decay electron. Alternatively, of the two
possibilities to reconstruct the $e\gamma$ invariant mass, the value closest to the $e^*$
candidate mass is chosen. The latter has been done for $m_{e^*} \geq 300$~GeV. Example mass
distributions for the two chosen options to reconstruct the $e^*$ candidate mass are shown in
Figs.~\ref{control2}(b) and (c). The alternatives of single-sided mass cuts and a mass window
are considered, leading to single-sided cuts for all values of $m_{e^*}$. Rejecting events
with both electrons or the photon in the EC leads to a slightly better sensitivity; since for
high values of $m_{e^*}$ the SM backgrounds are extremely small, we have not applied these
selection criteria for $m_{e^*} \geq 400$~GeV, in order to keep the search general beyond the
specific model considered here. Finally, the separation $\Delta {\cal R} (e_2,\gamma)$
between the lower $E_T$ electron and the photon allows discrimination between signal and
background for $m_{e^*} \leq 200$~GeV. This is illustrated in Fig.~\ref{control2}(d) for
$m_{e^*} = 100$~GeV. All mass-dependent selection criteria are summarized in Table
\ref{cutSummary}.

\begin{table}[hbt]
  \centering
  \caption{Mass-dependent selection criteria. The second and the third columns show the
           lower mass cuts. The next two columns show if events with both electrons or the
	   photon in the EC are kept, respectively, and in the last column the upper value
           for the separation between the second electron and the photon is given.}
  \begin{tabular}{cccccc}\hline\hline
  $m_{e^*}$ & $m(e_2,\gamma)$ & $m(e\gamma)^{\text{closest}}$ & EC/EC & EC       &  \\
  ~[GeV]~   & [GeV]           & [GeV]                      & $e$   & $\gamma$ & \rb{$\Delta {\cal R} (e_2,\gamma)$} \\\hline
  100	& $>90$  & --  		& no   & no   & $<1.8$ \\
  200	& $>165$ & --  		& no   & no   & $<3.3$ \\
  300	& --  	 & $>285$	& no   & no   & any    \\
  400	& --  	 & $>370$	& yes  & yes  & any    \\
  500	& --  	 & $>445$	& yes  & yes  & any    \\
  600	& --  	 & $>515$	& yes  & yes  & any    \\
  700	& --  	 & $>600$	& yes  & yes  & any    \\
  800	& --  	 & $>705$	& yes  & yes  & any    \\
  900	& --  	 & $>800$	& yes  & yes  & any    \\
  1000	& --  	 & $>900$	& yes  & yes  & any    \\\hline\hline
  \end{tabular}
  \label{cutSummary}
\end{table} 

\begin{table}
  \begin{center}
  \caption{For different values of the $e^*$ mass hypothesis, the number of selected data events, the SM expectation
           including statistical and systematic uncertainties, and the signal efficiency.}
  \begin{tabular}{ccr@{$\,\pm \,$}l@{$\,\pm \,$}ll@{$\,\pm \,$}l@{$\,\pm \,$}l}\hline\hline
    ~$m_{e^*}$~      &             & \multicolumn{3}{c}{ }                 & \multicolumn{3}{c}{Signal eff.} \\
    ~[GeV]~          & \rb{Data}   & \multicolumn{3}{c}{\rb{SM expectation}} & \multicolumn{3}{c}{[\%]}        \\ \hline
    100              &    0	   &  $0.33$    & $0.09$    & $0.03$       &  $13.2$ & $0.6$ & $1.3$   \\
    200              &    1	   &  $0.52$    & $0.16$    & $0.05$       &  $16.5$ & $0.6$ & $1.6$   \\
    300              &    1	   &  $0.32$    & $0.12$    & $0.03$       &  $22.2$ & $0.7$ & $2.2$   \\
    400              &    0	   &  $0.26$    & $0.11$    & $0.03$       &  $28.3$ & $0.8$ & $2.8$   \\
    500              &    0	   &  $0.12$    & $0.08$    & $0.01$       &  $31.5$ & $1.0$ & $3.1$   \\
    600              &    0	   & ~$(0.57$   & $0.54$    & $0.06)$  $\times 10^{-1}$~  &  $32.3$ & $0.9$ & $3.2$   \\
    700              &    0	   &  $(0.82$    & $0.37$ & $0.09)$    $\times 10^{-3}$   &  $34.3$ & $1.1$ & $3.4$   \\
    800              &    0	   &  $(0.48$    & $0.28$ & $0.06)$    $\times 10^{-3}$   & $32.2$ & $0.8$ & $3.2$   \\
    900              &    0	   &  $(0.17$    & $0.17$ & $0.02)$    $\times 10^{-3}$   & $33.2$ & $0.8$ & $3.3$   \\
    1000             &    0	   &  $(0.17$    & $0.17$ & $0.03)$    $\times 10^{-3}$   & $33.3$ & $0.9$ & $3.3$   \\\hline\hline
  \end{tabular}
  \label{results}
  \end{center}
\end{table}

The final selection efficiency varies from $13$\% ($m_{e^*} = 100$~GeV) up to $\approx
33$\% for higher values of $m_{e^*}$. In the data we find one event each for the
$m_{e^*} = 200$~GeV mass hypothesis and for the $m_{e^*} = 300$~GeV mass hypothesis,
respectively, and no events for other values of $m_{e^*}$, compatible with the SM
expectation. This result is summarized in Table \ref{results}.

The dominant systematic uncertainties are as follows. The uncertainty on the SM cross
sections is dominated by the DY process and the uncertainty from the choice of PDF and
renormalization and factorization scales [(3 -- 10)\%]. Electron reconstruction and
identification have an uncertainty of 2.5\% per electron, and a (1 -- 4)\% uncertainty
is assigned to the photon identification, depending on $\eta$ and $E_T$. The trigger
efficiency is $100^{+0}_{-3}$\%. The integrated luminosity is known to a precision of
$6.1$\% \cite{d0lumi}. The uncertainty on the number of background events due to jets
misidentified as photons is estimated to be 60\% of this particular background, from
differences between the expectation from the simulation and the independent measurement
from the data. A 25\% uncertainty is determined on the multijet background by comparing
the resulting multijet background estimate when using different criteria to select the
multijet background sample; after all selections, the multijet background is
negligible. The uncertainty on the signal cross section is estimated to be 10\%,
consisting of PDF uncertainties and missing higher order corrections.

\begin{figure} \setlength{\unitlength}{1cm}
  \begin{picture}(8.6,5.1)(0.0,0.0)
    \put(0.9,-0.3) {\includegraphics[scale=0.35]{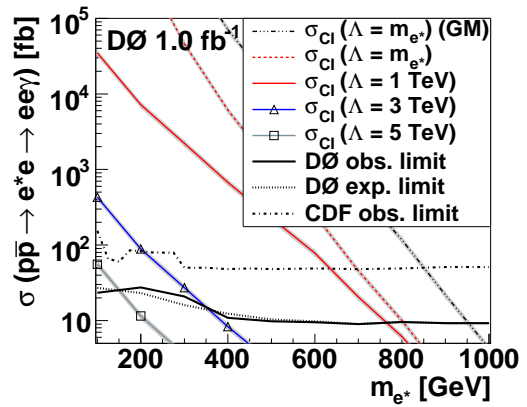}}
  \end{picture}
  \caption{The measured and expected limits on cross section times branching fraction,
           compared to the contact interaction model prediction for different choices
           of $\Lambda$. Also shown is the prediction under the assumption that no
           decays via contact interactions occur (``GM''), and the CDF result
           \cite{cdf}. The theoretical uncertainty of the model prediction is
	   indicated by shaded bands.}
  \label{limit1}
\end{figure}

\begin{figure} \setlength{\unitlength}{1cm}
  \begin{picture}(8.6,5.0)(0.0,0.0)
    \put(0.9,-0.3) {\includegraphics[scale=0.35]{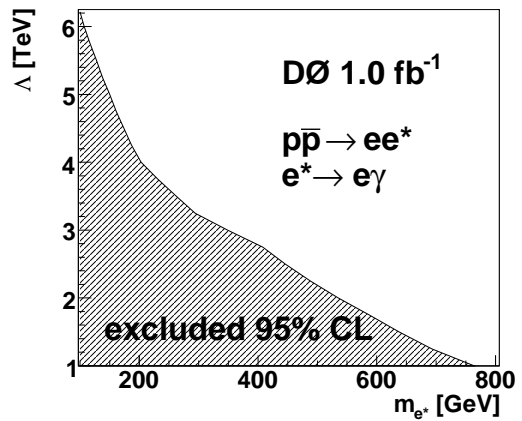}}
  \end{picture}
  \caption{The region in the $\Lambda$ -- $m_{e^*}$ plane excluded by the
  present analysis.}
  \label{limit2}
\end{figure}

Since the observed number of events is in agreement with the SM expectation, we set
95\% C.L.~limits on the $e^*$ production cross section times the branching fraction
into $e \gamma$. A Bayesian technique \cite{d0limit} is used, taking into account all
uncertainties and treating them as symmetric for simplicity. The resulting limit as a
function of $m_{e^*}$ is shown in Fig.~\ref{limit1} together with predictions of the
contact interaction model for different choices of the scale $\Lambda$. For $\Lambda =
1$~TeV ($\Lambda = m_{e^*}$), masses below 756~GeV (796~GeV) are excluded. In
Fig.~\ref{limit2}, the excluded region in terms of $\Lambda$ and $m_{e^*}$ is shown.

The CDF collaboration has recently searched \cite{cdf} for the production of excited
electrons using a data sample corresponding to an integrated luminosity of
$202$~pb$^{-1}$, but the CDF mass limit of $m_{e^*} > 879$~GeV at the 95\% C.L.~for
$\Lambda = m_{e^*}$ cannot be directly compared to ours for two reasons. The $e^*$
production cross section calculated with the version of {\sc pythia} used in
Ref.~\cite{cdf} is a factor of two higher than in subsequent versions corrected by the
{\sc pythia} authors. Furthermore, in Ref.~\cite{cdf}, it is assumed that decays via
contact interactions can be neglected, while in our analysis such decays are taken into
account in the calculation of the branching fraction $e^* \rightarrow e \gamma$,
following Ref.~\cite{Baur90}. Omitting contact interaction decays, we would obtain a
limit of $m_{e^*} > 946$~GeV for $\Lambda = m_{e^*}$ at the 95\%
C.L.\footnote{Multiplying the theoretical prediction in addition by a factor of two,
the mass limit would increase to 989~GeV.}.

In summary, we have searched for the production of excited electrons in the process $p
\bar p \rightarrow e^* e$ with $e^* \rightarrow e \gamma$, using about 1~fb$^{-1}$ of
data collected with the D0 detector. We find zero or one event in the data depending
on the mass of the hypothetical $e^*$, compatible with the SM expectation. We set
limits on the production cross section times branching fraction as a function of
$m_{e^*}$. For a scale parameter $\Lambda = 1$~TeV, masses below 756~GeV are excluded,
representing the most stringent limit to date.

%
We thank the staffs at Fermilab and collaborating institutions, 
and acknowledge support from the 
DOE and NSF (USA);
CEA and CNRS/IN2P3 (France);
FASI, Rosatom and RFBR (Russia);
CAPES, CNPq, FAPERJ, FAPESP and FUNDUNESP (Brazil);
DAE and DST (India);
Colciencias (Colombia);
CONACyT (Mexico);
KRF and KOSEF (Korea);
CONICET and UBACyT (Argentina);
FOM (The Netherlands);
Science and Technology Facilities Council (United Kingdom);
MSMT and GACR (Czech Republic);
CRC Program, CFI, NSERC and WestGrid Project (Canada);
BMBF and DFG (Germany);
SFI (Ireland);
The Swedish Research Council (Sweden);
CAS and CNSF (China);
and the
Alexander von Humboldt Foundation.
%


\begin{thebibliography}{99}
%
\bibitem[a]{alton}
Visitor from Augustana College, Sioux Falls, SD, USA.
\bibitem[b]{burdin}
Visitor from The University of Liverpool, Liverpool, UK.
\bibitem[c]{podesta-lerma}
Visitor from ICN-UNAM, Mexico City, Mexico.
\bibitem[d]{quadt,meyer}
Visitor from II. Physikalisches Institut, Georg-August-University, G{\"o}ttingen, Germany.
\bibitem[e]{voutilainen}
Visitor from Helsinki Institute of Physics, Helsinki, Finland.

\bibitem[\dag]{IntFellows}
Fermilab International Fellow.
\bibitem[\ddag]{deceased}
Deceased.

%
\vskip 0.25cm

  \bibitem{compositeness}
    H.~Terazawa, M.~Yasue, K.~Akama and M.~Hayashi, Phys.~Lett.~B {\bf 112}, 387 (1982);
    F.M.~Renard, Nuovo Cimento {\bf A77}, 1 (1983);
    A.~De Rujula, L. ~Maiani and R.~Petronzio, Phys.~Lett.~B {\bf 140}, 253 (1984);
    E.J.~Eichten, K.D.~Lane and M.E.~Peskin, Phys.~Rev.~Lett.~{\bf 50}, 811 (1983).

  \bibitem{Terazawa1} H.~Terazawa, Y.~Chikashige and K.~Akama, Phys.~Rev.~D {\bf 15}, 480 (1977);
                      Y.~Ne'eman, Phys.~Lett.~B {\bf 82}, 69 (1979).

  \bibitem{Baur90} U.~Baur, M.~Spira and P.M.~Zerwas, Phys.~Rev.~D {\bf 42}, 815 (1990).

  \bibitem{HERA} C.~Adloff {\it et al.} (H1 Collaboration), Phys.~Lett.~B {\bf 548}, 35 (2002);
         S.~Chekanov {\it et al.} (ZEUS Collaboration), Phys.~Lett.~B {\bf 549}, 32 (2002).

  \bibitem{LEP} G.~Abbiendi {\it et al.} (OPAL Collaboration), Phys.~Lett.~B {\bf 544}, 57 (2002);
         P.~Achard {\it et al.} (L3 Collaboration), Phys.~Lett.~B {\bf 568}, 23 (2003);
	 P.~Abreu {\it et al.} (DELPHI Collaboration), Eur.~Phys.~J.~C {\bf 8}, 41 (1999);
	 R.~Barate {\it et al.} (ALEPH Collaboration), Eur.~Phys.~J.~C {\bf 4},  571 (1998).

  \bibitem{comprun1} B.~Abbott {\it et al.} (D0 Collaboration), Phys.~Rev.~Lett.~{\bf 82}, 4769 (1999).

  \bibitem{cdf} D.~Acosta {\it et al.} (CDF Collaboration), Phys.~Rev.~Lett.~{\bf 94}, 101802 (2005).

  \bibitem{pythia} T.~Sj\"{o}strand {\it et al.}, Comput.~Phys.~Commun.~{\bf 135}, 238 (2001). {\sc Pythia}
                   version 6.3 is used throughout.

  \bibitem{dy} R.~Hamberg, W.L.~van Neerven and T.~Matsuura, Nucl.~Phys.~{\bf B359}, 343 (1991);
               Erratum Nucl.~Phys.~{\bf B644}, 403 (2002).

  \bibitem{mcfm} J.M.~Campbell and R.K.~Ellis, Phys.~Rev.~D {\bf 60}, 113006 (1999);
                 J.M.~Campbell and R.K.~Ellis, http://mcfm.fnal.gov/.

  \bibitem{ttbar} N.~Kidonakis and R.~Vogt, Int.~J.~Mod.~Phys.~A {\bf 20}, 3171 (2005).

  \bibitem{geant} R.~Brun and F.~Carminati, CERN Program Library Long Writeup W5013, 1994 (unpublished).

  \bibitem{cteq}  J.~Pumplin {\it et al.}, J.~High Energy Phys.~{\bf 0207}, 012 (2002);
                  D.~Stump {\it et al.}, J.~High Energy Phys.~{\bf 0310}, 046 (2003).

  \bibitem{run2det} V.~Abazov {\it et al.} (D0 Collaboration),
                    Nucl.~Instrum.~Methods Phys.~Res., Sect.~A {\bf 565}, 463 (2006).

  \bibitem{Zgamma} V.~Abazov {\it et al.} (D0 Collaboration), Phys.~Lett.~B {\bf 653}, 378 (2007).

  \bibitem{d0lumi} T.~Andeen {\it et al.}, Fermilab-TM-2365 (2007).

  \bibitem{d0limit} I.~Bertram {\it et al.} (D0 Collaboration), Fermilab-TM-2104 (2000).

\end{thebibliography}
\end{document}